\documentclass[prl,twocolumn,preprintnumbers,amsmath,amssymb]{revtex4}

\usepackage{graphicx}

\def\be{\begin{equation}}
\def\ee{\end{equation}}

\def\Tr{{\rm Tr}}
\def\HH{{\mathcal H}}

\def\a{\alpha}
\def\s{\sigma}

\def\bea{\begin{eqnarray}}
\def\eea{\end{eqnarray}}
\begin{document}

\title{Entanglement, combinatorics and finite-size effects in spin-chains}
\author{Bernard Nienhuis${}^{1}$, Massimo Campostrini${}^{2}$, and
  Pasquale Calabrese${}^{2,3}$} 
\affiliation{$^1$Institute for Theoretical Physics, Universiteit van Amsterdam,
1018 XE Amsterdam, The Netherlands,\\
$^{2}$INFN Pisa, Italy,
$^{3}$Dipartimento di Fisica dell'Universit\`a di Pisa, 
56127 Pisa, Italy}

\date{\today}

\begin{abstract}

  We carry out a systematic study of the exact block entanglement 
  in XXZ spin-chain at $\Delta=-1/2$.  
  We present, the first {\it analytic} expressions for reduced density matrices 
  of $n$ spins in a chain of length $L$ (for $n\leq 6$ and arbitrary but 
  odd $L$) of a truly interacting model.  
  The entanglement entropy, the moments of the reduced density
  matrix, and its spectrum are then easily derived.
  We explicitely construct the ``entanglement Hamiltonian'' as the logarithm
  of this matrix.
  Exploiting the degeneracy of the ground-state, we find the scaling 
  behavior of entanglement of the zero-temperature mixed state.

\end{abstract}

\maketitle

{\it Introduction}. Entanglement is a central concept in quantum information
science and it is becoming a common tool to study and analyze extended quantum 
systems because of its use in detecting the scaling behavior close to 
a quantum critical point \cite{rev}. It is has been pointed out that this 
scaling behavior is connected with the efficiency of 
numerical methods as quantum 
and density matrix
renormalization group (DMRG) \cite{rev}.
 
Let $\rho$ be the density matrix
of a system 
and let the Hilbert space be written as a direct product 
$\HH=\HH_A\otimes\HH_B$. 
$A$'s reduced density matrix is $\rho_A=\Tr_B \rho$
and the entanglement entropy is the corresponding Von Neumann entropy
\be
S_A=-\Tr_A \rho_A \log \rho_A\,,
\label{Sdef}
\ee
and analogously for $S_B$. 
When $\rho$ corresponds to a pure quantum state $S_A=S_B$.

When $A$ is a segment of length $n$ of an infinite one-dimensional
system in a critical ground state, the corresponding entanglement
entropy $S_{n}$  diverges as the logarithm of the
sub-system size \cite{Holzhey,Vidal,cc-04}
\be
S_n=\frac{c}3 \log n+ c'_1\,,
\label{SA}
\ee
where $c$ is the central charge of the associated conformal field theory (CFT)
and $c'_1$ a non-universal constant. 
Away from the critical point, $S_n$ saturates to a
constant \cite{Vidal} proportional to the logarithm of the correlation length
\cite{cc-04}. These properties made the entanglement entropy a basic 
tool to analyze 1D models. 
While it is impossible to mention here all the important
contributions in the field, we refer the interested reader to the
reviews \cite{rev}. 

In recent times, it has been remarked by few authors \cite{lh-08,cl-08} that 
the reduced density matrix $\rho_n$ contains much more
information than $S_n$. 
To the best of our knowledge the full reduced density matrix is known only for
free systems \cite{pesc} and is difficult to be obtained by numerical
methods because these tend to focus on its eigenvalues.
In order to go beyond free systems and to study the effect of
strong interactions, in this letter we report a first systematic study 
of the reduced density matrix of the antiferromagnetic XXZ chain at 
$\Delta=-1/2$ defined by the Hamiltonian
\be
H=-\sum_{j=1}^L [\s^x_j\s^x_{j+1}+\s^y_j \s^y_{j+1}+\Delta \s^z_j\s^z_{j+1}]\,,
\label{H0.5}
\ee 
with periodic boundary conditions ($\sigma_{L+1}=\sigma_1$) and an odd number of sites.
Here $\s^{x,y,z}_i$ stand for the Pauli matrices at the site $i$.
This critical model has the unique property that all the components of the 
ground-state wavefunction are integer 
multiples of the smallest one \cite{rs-01}. 
We will argue in the following that this property suffices
to get $\rho_n(L)$ for $n \leq 6$ and arbitrary $L$: 
an exceptional result for a truly interacting system.
Another unique feature of this spin-chain is that the ground-state 
energy, which is doubly
degenerate for any finite $L$ has no corrections to scaling:
$E_0=-(3/2) L$ exactly. 
We refer to the two ground states as $|\Psi_{\pm}\rangle$ 
with the upper index being the sign of the total spin in the $z$-direction.
 
This work is complementary to two recent papers. 
In Ref. \cite{ss-07} $\rho_n$ has been calculated in the thermodynamic (TD) 
limit (also for $n$ up to 6). 
In Ref. \cite{js-08} the connection with loop-models has been explored,
for a different measure of entanglement. 
In contrast, here the emphasis is on combinatorial and finite-size 
scaling aspects. 
The exact value of some elements of the reduced density matrix for 
smaller values of $n$ is also known for general $\Delta$ \cite{ger}
and all of them up to $n=6$ for $\Delta=-1$ \cite{sst-06}.

{{\it Analytic results for $\rho_n$}}.
Since the ground-state energy is exactly proportional to the system
size, and the Hamiltonian is represented by a matrix with rational
elements, also the ground-state vector has only rational components.
Suitably normalized, all ground-state components are integer
multiples of the smallest one. 
The ground-state for system sizes up to $L=25$ can be then obtained
with absolute precision in very modest computer time. 
With these ground-states we can construct the corresponding 
density matrices $\rho_n(L)$ with $n\leq L$.
Any element of  $\rho_n(L)$ is necessarily a rational number and in fact
a rational function of $L$, with numerator and denominator of degree
$\lceil  n^2/2\rceil$.
The data suffice to guess the denominator to be
 $2^{n^2} L^n \prod_{k=1}^{\lfloor n/2\rfloor}(L^2-4k^2)^{n-2k}$.
(Anti)symmetrized with respect to the two ground states  $\rho_n(L)$ turns out
to be an even (odd) function of $L$. As a result it 
can be determined completely for general $L$ and for $n\leq 6$.
For example, for $n=1$ and $n=2$ we obtained
\bea
&&\rho_1(L)=\frac1{2L}\left(
                   \begin{array}{cc}
                    L+1 & 0 \\
                    0 & L-1
                   \end{array}
                   \right),\\
&&\rho_2(L)=\frac{1}{2^4 L^2} \times\\
&&
\hskip -5mm
\left( \begin{array}{cccc}
2 (L+2)^2-2\!\!& 0 & 0 & 0 \\
0 &  6 L^2 - 6 & 5 L^2 + 3 &0\\
0 &  5 L^2 + 3 & 6 L^2 - 6 &0\\
0 &  0 & 0  & \!\!2(L-2)^2-2
      \end{array} \right).\nonumber
\eea
The other reduced density matrices are too large to display. 
We enclose an electronic Mathematica file (the density
matrix is {\tt rho[L,n]} with $L$ odd and $n=1,\cdots,6$).
In the mathematica file and in the following, the indeces of the matrix 
are the 
decimal form of the binary number representing the site product 
state ($1$ for $+$ and $0$ for $-$).

We observed the following properties 
of $\rho_n$:\\
(i) The first and the last elements correspond to the probability of a
string of equal spins, i.e. the emptiness formation
probabilities (EFP), 
$E_{\pm}(L,n)$, 
which is
\be
\prod_{k=0}^{n-1} \frac{
k!\;(3k+1)!\;(L-k-1)!\;(\frac{L\pm 1}{2}+k)!}{
(2k)!\;(2k+1)!\;(L+k)!\;(\frac{L\pm 1}{2}-k-1)!}
\label{emptiness}
\ee
for minority (see \cite{rs-01}) and majority spins respectively.
These two elements
approach  the same limit as $L\to\infty$, and are equal to
$\rho[1,1]=A_n/2^{n^2}$, where $A_n$ is the number of
$n\times n$ alternating sign matrices (ASM).\\
(ii) Some elements satisfy relations that connect
$\rho_n$ to $\rho_{n-1}$ when summing over one spin (see
the appendix A of Ref. \cite{ss-07}).
These equations can be used to derive expressions for some more
elements. For example, we have $\rho_{n+1}[1,1]+\rho_{n+1}[2,2]=\rho_n[1,1]$,
that combined with the ASM sequence for $\rho_n[1,1]$,
gives $\rho_n[2,2]=(2^n A_{n-1}-A_n)/2^{n^2}$. Similar relations can be
derived for a few other elements. \\ 
(iii) For $L\to\infty$ all non-zero elements remain non-zero 
and reproduce the results of Ref. \cite{ss-07}.\\
(iv) The analytic continuation to general $L$ satisfies 
$\rho_n(L)[i,j]=\rho_n(-L)[2^n+1-i,2^n+1-j]$.

\begin{figure}[t]
\includegraphics[width=8cm]{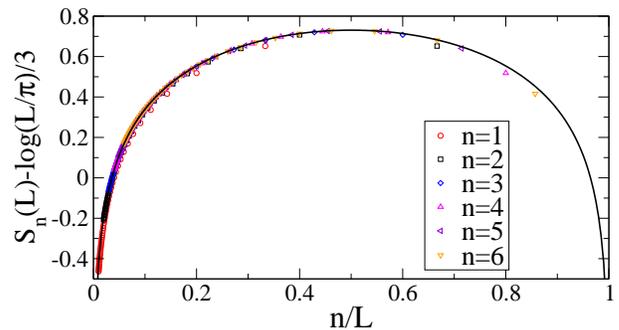} 
\caption{Finite size scaling of the entanglement entropy $S_n(L)$ against the
  CFT prediction $1/3 \log\sin (\pi n/L)+c'_1$ (full line). 
  We fixed $c_1'=0.7305$ \cite{c1p}.}
\label{Sn} 
\end{figure}

{\it The entanglement entropy}   
in the TD limit is 
\bea
S_1&=&\log 2,\quad S_2= 0.950749,\quad S_3=1.09287,\nonumber\\
S_4&=&1.19076, \quad S_5=1.26588, \quad S_6=1.32701
\eea 
the same as in Ref. \cite{ss-07}. 
We are in position to study the finite-size effects.
In Fig. \ref{Sn} we plot $S_n(L)-1/3 \log L/\pi$, and we compare it with the 
CFT prediction \cite{cc-04} $1/3 \log\sin (\pi n/L)+c'_1$ valid  for large 
enough $n$. We notice that
all the results fall on the same curve, except for small deviations at $n=1$. 
It is impressive and maybe unexpected that the asympotic scaling sets
in for such small value of $n$.

{\it Combinatorics}.
It has been noted in the study of Hamiltonian (\ref{H0.5}) 
that many physical quantities are sequences of integers or rationals 
that can be recognized in terms of known
ones, with the ASM sequence for the ground-state components that is the 
best known \cite{rs-01}, but not the only one \cite{rs-01,other,mn-04}.  

Once a sequence has been recognized, the critical 
exponents can be derived exactly from its asymptotic behavior, 
which in part motivates the great interest in this kind of studies. 
One could have hoped that the eigenvalues of the reduced density
matrices are rational or at least simple functions of $L$, but this is
not the case. 
It is then natural to move our attention to the elements of the density matrix
itself. Because of the conservation of the spin in the $z$ direction, the 
density matrix has non zero entries only between states with the same 
magnetization ($=-n/2,-n/2+1,\dots n/2-1,n/2$) and organizes into sectors, 
as evident from the block structure above. 
We already reported the formula for the first and last element of
$\rho_n(L)$, that are the only elements in the sectors with spin $\pm n/2$.
For the other sectors, the elements grow too quickly with $n$ to recognize
any simple behavior. For this reason we explored the possibility that some
non-trivial combinations of them could have a simpler structure.
Let us start by considering the sectors with total spin $\pm(n/2-1)$.
After some tries, we found that reasonable growing sequences are given 
by the trace of the matrix multiplied with the matrix $(-1)^{i+j}$. 
In the TD limit, this results in the following sequence:
\be 1,\; 2,\; 11,\; 72,\; 806,\; 14352, \ee
for both sectors, up to an overall factor $2^{-n^2}$.
This sequence grows with $n$ mildly, but we have been unable to recognize it. 
At finite $L$, the same trace for $\rho_n(2n+1)$ gives
\bea  
&&1,\; 2,\; 31,\; 124,\; 489,\; 1826,\;\;
6843,\; 25712,\; 97213, \nonumber \\ && \quad\; 369478,\;1410831,\; 5408272,\;\\ &&
\mbox{and}\;\; 2,\; 2,\; 2,\; 2,\; 2,\; 2,\; 2,\; 2,\cdots 
\eea
for the two sectors with $S_z=\pm(n/2-1)$ of $\rho_n(2n+1)$ respectively, 
up to the overall factor $E_-(2n+1,n)$ from Eq. (\ref{emptiness}).
The first sequence also grows in a reasonable way, but still we have not been
able to guess it. For second one, the recognition is instead obvious and 
provide a non-trivial relation whose physical origin is still unknown.
Similar sequences are easily derived for the other sectors.
We want to stress here that this generation of sequences is not only an
academic game. If we would have been able to find a sufficient number 
combinations of elements of the reduced density matrix that can be recognized,
we would have been access to analytic forms of the reduced density matrix for
any $n$ and $L$. By simply looking at the elements of the density matrix this
could seem an impossible task, but we explicitly showed that at least one
combination of them is very easy and that there are other sequences that do
not look prohibitive. The main reason why we reported these series here is to
stimulate further studies in this direction in order to eventually achieve the
knowledge of the full density matrix.

{\it The moments} of $\rho_n(L)$
\be
R_n^{(\a)}(L)= \Tr\, \rho_n^\a(L)\,,
\ee
for $\a$ integer are sequences of rationals. 
For critical systems they display the universal asymptotic behavior 
\cite{Holzhey,cc-04} 
\be
R_n^{(\a)}(L)=c_\a \left[\frac{L}\pi \sin \frac{\pi
    n}L\right]^{-c(\a-1/\a)/6}\,,
\label{RCFT}
\ee
from which one can reconstruct the full spectrum of $\rho_n(L)$ \cite{cl-08}
and its behavior is connected with the accuracy of some numerical
algorithms based on matrix product states \cite{num}. 
Despite this universal behavior $R_n^{(\a)}$ has been only marginally
considered in the literature \cite{cc-04,cum}.
In the TD limit, since all elements of $\rho_n(\infty)$ have a common
denominator $2^{n^2}$, all moments can be written as 
\be 
R^{(\a)}_n= {r_n^{(\a)}}\,{2^{-\a n^2}},\quad {\rm with}\; r_n^{(\a)}\;
{\rm integers}.
\ee 
For example, the values of $r_n^{(2)}$ up to $n=6$ are 
\bea
&&r^{(2)}_1=2,\quad r^{(2)}_2=130,\quad r^{(2)}_3=107468,\nonumber\\
&&r^{(2)}_4=1796678230,\quad r^{(2)}_5=413605561988912,\nonumber\\
&&r^{(2)}_6=1768256505302499935380\,.
\eea
We are not able to recognize this fast growing sequence.
Increasing $\a$, the growth is even faster.

The numerical values of $R^{(2)}_n$:
\bea
&&R_1=0.5,\quad R_2=0.507813,\quad R_3=0.409958,\qquad\\
&&R_4=0.418322,\quad R_5=0.367356,\quad R_6=0.374443.\nonumber
\eea
clearly display even-odd oscillations that prevent us from any precise
scaling analysis as the previous one for the entropy.
Multiplying this by $n^{1/4}$ (see
Eq. (\ref{RCFT}))  we get 
\be
0.5,\quad 0.604,\quad 0.540, \quad0.592, \quad0.549, \quad 0.586, 
\ee
that tend to approach a constant value confirming the CFT scaling.
A systematic study of these oscillations requires the analysis of larger
values of $n$, that are not accessible to the present method and for which we
performed DMRG calculations that appear
elsewhere \cite{ccn-prep}.
For periodic boundary conditions, 
these oscillations are only present for $\a\neq1$, and not  for 
the entanglement entropy. Thus they have a different origin than those
found elsewhere for $S_n$ that are due to boundary 
effects \cite{lsca-06,dmcf-06}. 
The same kind of oscillations have been found numerically 
for the multi-interval entanglement \cite{fps-09}. 
It is rather natural that all the elements of $\rho_n(L)$ 
oscillate as a consequence of the tendency to antiferromagnetic 
order of the XXZ chain at $\Delta=-1/2$, and consequently most of the 
averages that are calculated from them are expected to oscillate as the 
moments do. Why and how these oscillations cancel between each other only for 
the von Neumann entropy is still mysterious.

\begin{figure}[t]
{\includegraphics[width=8cm]{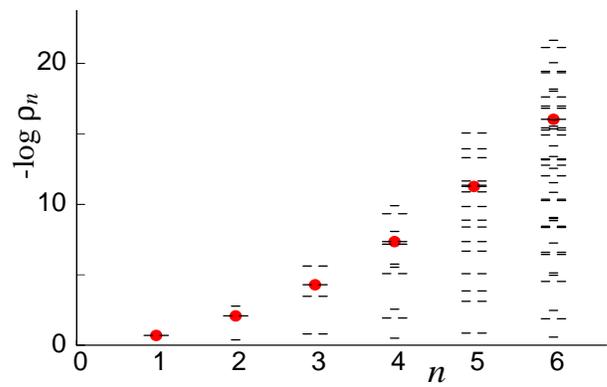} }
\caption{Spectrum of $\rho_n$ in the TD limit, the dots represent the
  first and last element Eq. (\ref{emptiness}).
Note that these dots sit at $3/4$ ($2/3$) of the spectrum for odd (even) $n$.
}
\label{spec} 
\end{figure}

$R_n^{(\a)}$ is not the only scaling quantity
that can be represented as sequence of rationals.  
A good alternative is given by the central values
$Q_n^{(\a)}\equiv R^{(\a)}_n (2n+1)$, which grows more slowly.
In fact, in the matrix $\rho_n(2n+1)$ the common denominator is 
the inverse of $E_-(2n+1,n)$ in Eq. (\ref{emptiness}).
As a typical example we report $Q_n^{(2)}=q_n E^2_-(2n+1,n)$: 
\bea
 q_1=5,\;\; q_2=327,\;\; q_3=159502, \;\;q_4=680263760,\nonumber \\
q_5=22821555833635
,\;\; 
q_6=6408136183930928388\,.
 \eea
Unfortunately, although it grows slower than $r_n^{(2)}$,  we are also 
unable to recognize the sequence, but the guessing seems less prohibitive.


{\it The spectrum} of $\rho_n$ in the TD limit is shown in Fig. 
\ref{spec}. Several interesting features are evident in the plot. 
For odd $n$ all the eigenvalues are doubly degenerate, while for even $n$ only
some. The spectrum at $n$ is roughly 
repeated at the bottom and at the top of the spectrum at $n+2$ with some
new structure in between. This is also what happens for free fermions \cite{pesc} ($\Delta=0$ in Eq. (\ref{H0.5})). 
Thus the interaction appears to
change only quantitative features of the spectrum and not the qualitative ones. 
This is highly non-trivial because non-zero $\Delta$ introduces
well-known strong non-perturbative effects.
The smallest eigenvalue scales like ${\rm const}^{n^2}$ (i.e. the top of Fig. \ref{spec}
is almost a parabola). Such scaling is known to be true for the 
``all up'' eigenvalue $\rho[1,1]$, i.e. to the EFP \cite{mn-04}, 
that is marked as a dot in the figure. 

\begin{figure}[t]
\includegraphics[width=8cm]{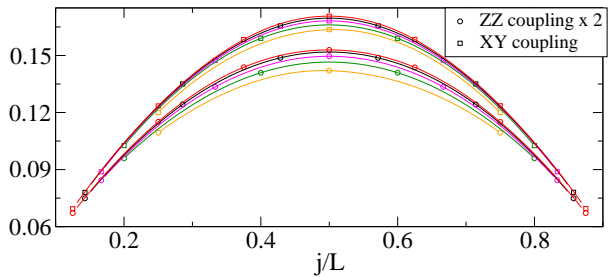} 
\caption{Dependence of the coupling constants of the entanglement Hamiltonian 
on the position in the block. Squares (circles) refer to the XY (2 times ZZ) 
couplings, while the full lines are only guides for the eyes to connect points 
at the same $n$.}
\label{Ham} 
\end{figure}

{\it The logarithm of the density matrix} can be interpreted as an effective 
Hamiltonian for the subsystem through the natural definition 
$\rho_n(L)=e^{-\hat{H}_n(L)}$ at a fictitious temperature $T=1$ (as observed
independently \cite{lh-08}). 
This effective Hamiltonian can be written down exactly only for free fermions
\cite{pesc},
but it is extremely difficult to obtain even part of it for
interacting systems, because it requires the knowledge of the full density
matrix. 
The present study gives this unique opportunity (it is enough to diagonalize
the density matrix, taking the logarithm of the eigenvalues and then use the
diagonalizing transformation to go back to the original spin basis).
In the TD limit we find that the largest terms are a diagonal interaction, 
$J^z_jS^z_jS^z_{j+1}$, and a
nearest neighbor exchange $J_j S^+_jS^-_{j+1}$+hc with a ratio 
$\sim -0.55$ that is almost the same as in the original model $-0.5$.
All other terms (exchanges over larger distance, exchange of more than
one pair of spins, multispin interaction) are more than one order of
magnitude smaller. 
The couplings depend on the position roughly quadratically
\be
J_j^z(n)\simeq A \frac{j(n-j)}n\,,
\ee
with $A \sim 0.6$. The parabolic dependence of the coupling constants is
shown in Fig. \ref{Ham}.

\begin{figure}[t]
\includegraphics[width=8cm]{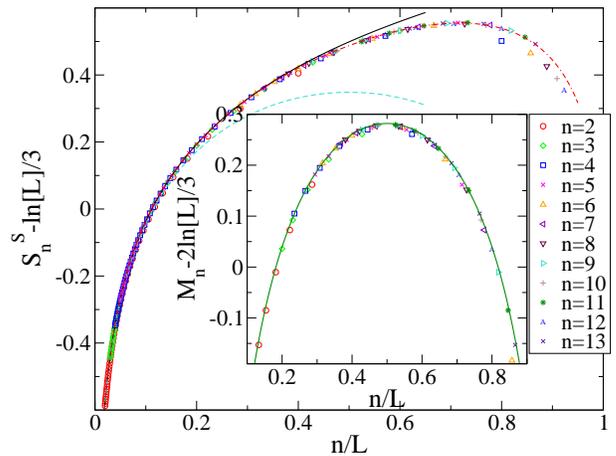} 
\caption{Entanglement entropy of the symmetrized density matrix 
$S_n^s(L)$ versus $n/L$ and its asymptotic behavior.
The full line is $(\log n)/3 +c'_1$ that is an effective description up to 
$n/L\sim 0.5$. The dashed line is the finite-size CFT prediction for pure state
$(\log L/\pi \sin \pi n/L)/3 +c'_1$ that clearly does not work for $x\geq
0.2$.
The dashed-dotted curve is the heuristic formula  
$S_n^{(s)}(L=n/x)\simeq 1/3 \log\left[n(1-A \sinh^4 x)\right]$ with 
$A\simeq 0.48$ that works well for all values of $n$ (except very small $n$ 
and $L-n$).
Inset: Scaling of the mutual information $M_n$ as function of $n/L$ compared
with the symmetrized heuristic guess.}
\label{eeex} 
\end{figure}

{\it The symmetrized density matrix}.
The degeneracy of the ground-state at finite $L$ gives another unique
opportunity: the symmetrized density matrix
\be
\rho^s=\frac12(|\Psi_+\rangle\langle \Psi_+|+
|\Psi_-\rangle\langle \Psi_-|)\,,
\ee
corresponds to minimum energy and it is a zero temperature mixed state.
$\rho^s$ has no interpretation in CFT, and so it is a new quantity. 
In the TD limit $S(\rho^s_n)=S(\rho_n)$. 
In Fig. \ref{eeex} we report the exact results for the 
entanglement entropy $S^s_n(L)$ of this mixed state. Up to $n/L\sim 0.5$
they are well described by
$S^s_n(L)=(\log n)/3 +c'_1$,
({\em independent} of $L$ and 
with the same $c'_1$ as before) as shown by the full line in the plot. 
Note that a finite size correction in the form of a sine (dashed-line) does not 
work for $x>0.2$. 
However, a good collapse is observed for all $x$. (the last points 
that do not fall on the master curve  correspond to $n=L-1$ and are not 
expected to scale). 
We find that the corrections to the  $\log n$ behavior are
of the order of $(n/L)^4$. 
The collapsed data are fit remarkably 
well by
$S_n^{(s)}(L=n/x)\simeq 1/3 \log\left[n(1-A \sinh^4 x)\right]$ with $A\simeq 0.48$.
We do not claim that this form has any justification.

In a mixed state the entropy (\ref{Sdef}) is not a good measure of
entanglement because it mixes classical and quantum correlations. From
quantum information we know that an appropriate measure is the mutual
information
\be
M_n=S_n+S_{L-n}-S_L\,.
\ee
The calculation of $M_n$ is more difficult because even for small $n$
we need $S_{L-n}$ that can correspond to a large block. 
The data available from the exact ground state (up to $n=13$, $L=21$ plotted 
in the inset of Fig. \ref{eeex}) collapse and define a 
universal function that is described by the symmetrized version of the
heuristic $S^{(s)}_n(L)$ introduced before,
plotted as a full line in Fig. \ref{eeex}.

Such mixed zero-temperature states are present every time that the 
ground-state is degenerate at finite $L$. Among these, supersymmetric lattice 
models \cite{susy} are very interesting and 
they could be understood more deeply in this framework.

{\it Conclusions and perspective}.  
We presented explicit analytic expressions for the reduced density matrix
of the XXZ spin-chain at $\Delta=-1/2$.  From these matrices we built
several sequences of integers that encode the scaling behavior. From 
the exact density matrix, we constructed the entanglement
Hamiltonian, a result that is not easily accessible to other approaches.
We found the remarkable property that this Hamiltonian is dominated 
by nearest-neighbor terms of the same form as the original Hamiltonian.
Finally, because the ground-state is doubly degenerate, we could study the 
entanglement of a zero-temperature mixed state, showing a very different 
behavior from a pure one. 

These results are unique in two ways: they concern the full density
matrix rather than only its entropy, and they are completely exact
rather than asymptotic.  
They suggest many questions that can be posed also numerically for 
more general systems. 
They show explicitly that even for modest block size, asymptotic 
behavior is evident, eventually allowing for
the study of further finite size corrections. 
Finally, since the ground state is known to have strong connections 
with combinatorial problems of current interest \cite{rs-01,js-08,other,mn-04}, 
it is expected that further properties of the
density matrix for arbitrarily large blocks can be found by
combinatorial methods.

{\it Acknowledgments}. PC benefited of a travel grant from ESF (INSTANS
activity).


\begin{thebibliography}{99}
\bibitem{c1p}
A precise determination of the non-universal constants $c_1'$ and $c_\a$
has been obtained via DMRG \cite{ccn-prep}.
\end{thebibliography}
\end{document}